# Limited Lifespan of Fragile Regions in Mammalian Evolution

Max A. Alekseyev[1] and Pavel A. Pevzner[2]

[1] University of South Carolina, Columbia, SC, U.S.A.
[2] University of California, San Diego, CA, U.S.A.

**Abstract.** An important question in genome evolution is whether there exist fragile regions (*rearrangement hotspots*) where chromosomal rearrangements are happening over and over again. Although nearly all recent studies supported the existence of fragile regions in mammalian genomes, the most comprehensive phylogenomic study of mammals (Ma et al. (2006) *Genome Research* **16**, 1557-1565) raised some doubts about their existence. We demonstrate that fragile regions are subject to a "birth and death" process, implying that fragility has limited evolutionary lifespan. This finding implies that fragile regions migrate to different locations in different mammals, explaining why there exist only a few chromosomal breakpoints shared between different lineages. The birth and death of fragile regions phenomenon reinforces the hypothesis that rearrangements are promoted by matching segmental duplications and suggests putative locations of the currently active fragile regions in the human genome.

## Introduction

In 1970 Susumu Ohno [35] came up with the Random Breakage Model (RBM) of chromosome evolution, implying that there are no rearrangement hotspots in mammalian genomes. In 1984 Nadeau and Taylor [34] laid the statistical foundations of RBM and demonstrated that it was consistent with the human and mouse chromosomal architectures. In the next two decades, numerous studies with progressively increasing resolution made RBM the *de facto* theory of chromosome evolution.

RBM was refuted by Pevzner and Tesler, 2003 [38] who suggested the Fragile Breakage Model (FBM) postulating that mammalian genomes are mosaics of fragile and solid regions. In contrast to RBM, FBM postulates that rearrangements are mainly happening in fragile regions forming only a small portion of the mammalian genomes. While the rebuttal of RBM caused a controversy [7, 43, 44], Peng et al., 2006 [36] and Alekseyev and Pevzner, 2007 [2] revealed some flaws in the arguments against FBM. Furthermore, the rebuttal of RBM was followed by many studies supporting FBM [6, 8, 9, 12, 14, 17, 19, 20, 21, 24, 27, 28, 29, 30, 31, 39, 40, 41, 46, 47, 49, 51].

Comparative analysis of the human chromosomes reveals many short adjacent regions corresponding to parts of several mouse chromosomes [16]. While such a surprising arrangement of synteny blocks points to potential rearrangement hotspots, it remains unclear whether these regions reflect genome rearrangements or duplications/assembly errors/alignment artifacts. Early studies of genomic architectures were unable to distinguish short synteny blocks from artifacts and thus were limited to constructing large synteny blocks. Ma et al., 2006 [25] addressed the challenge of constructing high-resolution synteny blocks via the analysis of multiple genomes. Remarkably, their analysis suggests that there is limited breakpoint reuse, an argument against FBM, that led to a split among researchers studying chromosome evolution and raised a challenge of reconciling these contradictory results. Ma et al., 2006 [25] wrote: *"a careful analysis* [of the RBM vs FBM controversy] *is beyond the scope of this study"* leaving the question of interpreting their findings open.

Various models of chromosome evolution imply various statistics and thus can be verified by various tests. For example, RBM implies exponential distribution of the synteny block sizes, consistent with the human-mouse synteny blocks observed in [34]. Pevzner and Tesler, 2003 [38] introduced the "pairwise breakpoint reuse" test and demonstrated that while RBM implies low breakpoint reuse, the human-mouse synteny blocks expose rampant breakpoint reuse. Thus RBM is consistent with the "exponential length distribution" test [34] but inconsistent with the "pairwise breakpoint reuse" test [37]. Both these tests are applied to *pairs* of genomes, not taking an advantage of multiple genomes that were recently sequenced. Below we introduce the "multispecies breakpoint reuse" test and demonstrate that both RBM and FBM do not pass this test. We further propose the *Turnover Fragile Breakage Model (TFBM)* that extends FBM and complies with the multispecies breakpoint reuse test.

Technically, findings in [25] (limited breakpoint reuse between different lineages) are not in conflict with findings in [38] (rampant breakpoint reuse in chromosome evolution). Indeed, Ma et al., 2006 [25] only considered *inter-reuse* between different branches of the phylogenetic tree and did not analyze *intra-reuse* within individual branches of the tree. TFBM reconciles the recent studies supporting FBM with the Ma et al., 2006 [25] analysis. We demonstrate that data in [25] reveal rampant but elusive breakpoint reuse that cannot be detected via counting repeated breakages between various pairs of branches of the evolutionary tree. TFBM is an extension of FBM that reconciles seemingly contradictory results in [6, 8, 9, 12, 14, 17, 19, 20, 21, 24, 27, 28, 29, 30, 31, 39, 40, 41, 46, 47, 49, 51] and [25] and explains that they do not contradict to each other. TFBM postulates that fragile regions have a limited lifespan and implies that they can migrate between different genomic locations. The intriguing implication of TFBM is that few regions in a genome are fragile at any given time raising a question of finding the currently active fragile regions in the human genome.

While many authors have discussed the causes of fragility, the question what makes certain regions fragile remains open. Previous studies attributed fragile regions to segmental duplications [5, 18, 42, 50], high repeat density [32], high recombination rate [33], pairs of tRNA genes [10, 23], inhomogeneity of gene distribution [36], and long regulatory regions [17, 30, 36]. Since we observed the birth and death of fragile regions, we are particularly interested in features that are also subject to birth and death process. Recently, Zhao and Bourque, 2009 [50] provided a new insight into association of rearrangements with segmental duplications by demonstrating that many rearrangements are flanked by *Matching Segmental Duplications* (MSDs), i.e., a pair of long similar regions located within a pair of breakpoint regions corresponding to a rearrangement event. MSDs arguably represent an ideal match for TFBM among the features that were previously implicated in breakpoint reuses. TFBM is consistent with the hypothesis that MSDs promote fragility since the similarity between MSDs deteriorates with time, implying that MSDs are also subjects to a "birth and death" process.

## Results

### Rearrangements and breakpoint graphs

For the sake of simplicity, we start our analysis with *circular genomes* consisting of circular chromosomes. While we use circular chromosomes to simplify the computational concepts discussed in the paper, all analysis is done with real (linear) mammalian chromosomes (see Alekseyev, 2008 [1] for subtle differences between circular and linear chromosome analysis). We represent a circular chromosome with synteny blocks $x_1, \ldots, x_n$ as a cycle (Fig. 1) composed of $n$ directed labelled edges (corresponding to the blocks) and $n$ undirected unlabeled edges (connecting adjacent blocks). The directions of the edges correspond to *signs* (strands) of the blocks. We label the *tail* and *head* of a directed edge $x_i$ as $x_i^t$ and $x_i^h$ respectively. We represent a genome as a *genome graph* consisting of disjoint cycles (one for each chromosomes). The edges in each cycle alternate between two colors: one color reserved for undirected edges and the other color (traditionally called "obverse") reserved for directed edges.

Let $P$ be a genome represented as a collection of *alternating* black-obverse cycles (a cycle is alternating if the colors of its edges alternate). For any two black edges $(u, v)$ and $(x, y)$ in the genome (graph) $P$ we define a *2-break* rearrangement (see [3]) as replacement of these edges with either a pair of edges $(u, x), (v, y)$, or a pair of edges $(u, y), (v, x)$ (Fig. 2). 2-breaks extend the standard operations of reversals (Fig. 2a), fissions (Fig. 2b), or fusions/translocations (Fig. 2c) to the case of circular chromosomes. We say that a 2-break on edges $(u, x), (v, y)$ *uses* vertices $u, x, v$ and $y$.

Let $P$ and $Q$ be "black" and "red" genomes on the same set of synteny blocks $\mathcal{X}$. The *breakpoint graph* $G(P, Q)$ is defined on the set of vertices $V = \{x^t, x^h \mid x \in \mathcal{X}\}$ with black and red edges inherited from genomes $P$ and $Q$ (Fig. 1). The black and red edges form a collection of alternating *black-red cycles* in $G(P, Q)$ and play an important role in analyzing rearrangements (see [11] for background information on genome rearrangements). The *trivial cycles* in $G(P, Q)$, formed by pairs of parallel black and red edges, represent common adjacencies between synteny blocks in genomes $P$ and $Q$. Vertices of the non-trivial cycles in $G(P, Q)$ represent *breakpoints* that partition genomes $P$ and $Q$ into $(P, Q)$-synteny blocks.

The 2-*break distance d(P, Q)* between circular genomes *P* and *Q* is defined as the minimum number of 2-breaks required to transform one genome into the other. In contrast to the genomic distance [13] (for linear genomes), the 2-break distance for circular genomes is easy to compute [48]:

**Theorem 1** *The* 2-*break distance between circular genomes P and Q is $b(P,Q) - c(P,Q)$ where $b(P,Q)$ and $c(P,Q)$ are respectively the number of $(P,Q)$-synteny blocks and non-trivial black-red cycles in $G(P,Q)$.*

**Inter- and intra- breakpoint reuse**

Figure 3 shows a phylogenetic tree with specified rearrangements on its branches (we write $\rho \in e$ to refer to a 2-break $\rho$ on an edge $e$). We represent each genome as a genome graph (i.e., a collection of cycles) on the same set $V$ of $2n$ vertices (corresponding to the endpoints of the synteny blocks). Given a set of genomes and a phylogenetic tree describing rearrangements between these genomes, we define the notions of inter- and intra-breakpoint reuses. A vertex $v \in V$ is *inter-reused* on two distinct branches $e_1$ and $e_2$ of a phylogenetic tree if there exist 2-breaks $\rho_1 \in e_1$ and $\rho_2 \in e_2$ that both use $v$. Similarly, a vertex $v \in V$ is *intra-reused* on a branch $e$ if there exist two distinct 2-breaks $\rho_1, \rho_2 \in e$ that both use $v$. For example, a vertex $c^h$ is inter-reused on the branches $(Q_3, P_1)$ and $(Q_2, P_3)$, while a vertex $f^h$ is intra-reused on the branch $(Q_3, Q_2)$ of the tree in Fig. 3. We define $br(e_1, e_2)$ as the number of vertices inter-reused on the branches $e_1$ and $e_2$, and $br(e)$ as the number of vertices intra-reused on the branch $e$. An alternative approach to measuring breakpoint intra-reuse is to define *weighted intra-reuse* of a vertex $v$ on a branch $e$ as $\max\{0, use(e,v) - 1\}$ where $use(e,v)$ is the number of 2-breaks on $e$ using $v$. The weighted intra-reuse $BR(e)$ on the branch $e$ is the sum of weighted intra-reuse of all vertices.[3]

Given simulated data, one can compute $br(e)$ for all branches and $br(e_1, e_2)$ for all pairs of branches in the phylogenetic tree. However, for real data, rearrangements along the branches are unknown, calling for alternative ways for estimating the inter- and intra-reuse.

Cycles in the breakpoint graphs provide yet another way to estimate the inter- and intra-reuse. For a branch $e = (P, Q)$ of the phylogenetic tree, one can estimate $br(e)$ by comparing the reversal distance $d(P, Q)$ and the number of breakpoints $b(P, Q)$ between the genomes $P$ and $Q$. It results in the lower bound $bound(e) = 4 \cdot d(P, Q) - 2 \cdot b(P, Q)$ for $BR(e)$ [37] that also gives a good approximation for $br(e)$. On the other hand, one can estimate $br(e_1, e_2)$ as the number $bound(e_1, e_2)$ of vertices shared between non-trivial cycles in the breakpoint graphs corresponding to the branches $e_1$ and $e_2$ (similar approach was used in [22] and later explored in [25, 31]). Assuming that the genomes at the internal nodes of the phylogenetic tree can be reliably reconstructed [4, 25, 26, 45], one can compute $bound(e)$ and $bound(e_1, e_2)$ for all (pairs of) branches. Below we show that these bounds accurately approximate the intra- and inter-reuse.

---

[3] We remark that if no vertex is used more than twice on a branch $e$ then $BR(e) = br(e)$.

**Analyzing breakpoint reuse (simulated genomes)**

We start from analyzing simulated data based on FBM with $n$ fragile regions present in $k$ genomes that evolved according to a certain phylogenetic tree (for the varying parameter $n$). We represent one of the leaf genomes as the genome with 20 random circular chromosomes and simulate hundred 2-breaks on each branch of the tree.

Figure 4 represents a phylogenetic tree on five leaf genomes, denoted $M, R, D, Q, H$, and three ancestral genomes, denoted $MR, MRD, QH$. Table 1(left panel) presents the results of a single FBM simulation and illustrates that $bound(e_1, e_2)$ provides an excellent approximation for inter-reuses $br(e_1, e_2)$ for all 21 pairs of branches.[4] While $bound(e)$ (on the diagonal of Table 1, left panel) is somewhat less accurate, it also provides a reasonable approximation for $br(e)$.

Below we describe analytical approximations for the values in Table 1(left panel). Since every 2-break uses 4 out of $2n$ vertices in the genome graph, a random 2-break uses a vertex $v$ with the probability $\frac{2}{n}$. Thus, a sequence of $t$ random 2-breaks does not use a vertex $v$ with the probability $(1 - \frac{2}{n})^t \approx e^{-\frac{2t}{n}}$ (for $t \ll n$). For branches $e_1$ and $e_2$ with respectively $t_1$ and $t_2$ random 2-breaks, the probability that a particular vertex is inter-reused on $e_1$ and $e_2$ is approximated as $(1 - e^{-\frac{2t_1}{n}}) \cdot (1 - e^{-\frac{2t_2}{n}})$. Therefore, the expected number of inter-reused vertices is approximated as $2n \cdot (1 - e^{-\frac{2t_1}{n}}) \cdot (1 - e^{-\frac{2t_2}{n}})$. Below we will compare the observed inter-reuse with the expected inter-reuse in FBM to see whether they are similar thus checking whether FBM represents a reasonable null hypothesis. We will use the term *scaled inter-reuse* to refer to the observed inter-reuse divided by the expected inter-reuse. If FBM is an adequate null hypothesis we expect the scaled inter-reuse to be close to 1.

Similarly, a sequence of $t$ random 2-breaks uses a vertex $v$ exactly once with the probability $t \cdot \frac{2}{n} \cdot \left(1 - \frac{2}{n}\right)^{t-1} \approx \frac{2t}{n} e^{\frac{2(t-1)}{n}}$. Therefore, the probability of a particular vertex being intra-reused on a branch with $t$ random 2-breaks is approximately $1 - e^{-\frac{2t}{n}} - \frac{2t}{n} e^{\frac{2(t-1)}{n}}$, implying that the expected intra-reuse is approximately $2n \cdot \left(1 - e^{-\frac{2t}{n}} - \frac{2t}{n} e^{\frac{2(t-1)}{n}}\right)$. We will use the term *scaled intra-reuse* to refer to the observed intra-reuse divided by the expected intra-reuse. Our simulations showed that the scaled intra- and inter-reuse for 21 pairs of branches are all close to 1 (data are not shown).

We also performed a similar simulation, this time varying the number of 2-breaks on the branches according to the branch lengths specified in Fig. 4. Again, the lower bounds provide accurate approximations in the case of varying branch lengths. Similar results were obtained in the case of evolutionary trees with varying topologies (data are not shown). We therefore use only lower bounds to generate Table 1(right panel) rather than showing both real distances and the lower bounds as in Table 1(left panel).

---

[4] We remark that $bound(e_1, e_2) = br(e_1, e_2)$ if simulations produce the shortest rearrangement scenarios on the branches $e_1$ and $e_2$. Table 1(left panel) illustrates that this is mainly the case for our simulations.

In the case when the branch lengths vary, we find it convenient to represent data as a plot that illustrates variability in the scaled inter-use. We define the *distance* between branches $e_1$ and $e_2$ in the phylogenetic tree as the distance between their midpoints, i.e., the overall length of the path, starting at $e_1$ and ending at $e_2$, minus $\frac{d(e_1)+d(e_2)}{2}$. For example, $d(M+, H+) = 56+170+58+28-\frac{56+28}{2} = 270$ (see Fig. 4). The $x$-axis in Fig. 5 represents the distances between pairs of branches (21 pairs total), while $y$-axis represents the scaled inter-reuse for pairs of branches at the distance $x$.

**Surprising irregularities in breakpoint reuse in mammalian genomes**

The branch lengths shown in Fig. 4 actually represent the approximate numbers of rearrangements on the branches of the phylogenetic tree for *M*ouse, *R*at, *D*og, maca*Q*ue, and *H*uman genomes (represented in the alphabet of 433 "large" synteny blocks exceeding 500,000 nucleotides in human genome [4]). For the mammalian genomes, $M, R, D, Q$, and $H$, we first used MGRA [4] to reconstruct genomes of their common ancestors (denoted *MR*, *MRD*, and *QH* in Fig. 4) and further estimated the breakpoint inter-reuse between pairs of branches of the phylogenetic tree. The resulting Table 2 reveals some striking differences from the simulated data (Table 1, right panel) that follow a peculiar pattern: the larger is the distance between two branches, the smaller is the amount of inter-reuse between them (in contrast to RBM/FBM where the amount of inter-reuse does not depend on the distance between branches). The statement above is imprecise since we haven't described yet how to compare the amount of inter-reuse for different branches at various distances. However, we can already illustrate this phenomenon by considering branches of similar length that presumably influence the inter-reuse in a similar way (see below).

We notice that branches *M+*, *R+*, and *QH+* have similar lengths (varying from 56 to 68 rearrangements) and construct subtables of Table 1(right panel) (for $n$ = 900) and Table 2 with only three rows corresponding to these branches (Table 3). Since the lengths of branches *M+*, *R+*, and *QH+* are similar, FBM implies that the elements belonging to the same columns in Table 3 should be similar. This is indeed the case for simulated data (small variations within each column) but not the case for real data. In fact, maximal elements in each column for real data exceed other elements by a factor of 3-5 (with an exception of the *MR+* column). Moreover, the peculiar pattern associated with these maximal elements (maximal elements correspond to red cells) suggests that this effect is unlikely to be caused by random variations in breakpoint reuses. We remind the reader that red cells correspond to pairs of adjacent branches in the evolutionary tree suggesting that breakpoint reuse is maximal between close branches and is reducing with evolutionary time. A similar pattern is observed for the other pairs of branches of similar length: adjacent branches feature much higher inter-reuse than distant branches. We also remark that the most distant pairs of branches (*H+* and *M+*, *H+* and *R+*, *Q+* and *M+*, *Q+* and *R+* in the yellow cells) feature the lowest inter-reuse. The only branch that shows relatively similar inter-reuse (varying from 58 to 80) with the branches *M+*, *R+*, and *QH+* is the branch *MR+* which is adjacent to each of these branches.

Below we modify FBM to come up with a new model of chromosome evolution, explaining the surprising irregularities in the inter-reuse across mammalian genomes.

**Turnover Fragile Breakage Model: birth and death of fragile regions**

We start with a simulation of 100 rearrangements on every branch of the tree in Fig. 4. However, instead of assuming that fragile regions are fixed, we assume that after every rearrangement $x$ fragile regions "die" and $x$ fragile regions are "born" (keeping a constant number of fragile regions throughout the simulation). We assume that the genome has $m$ potentially "breakable" sites but only $n$ of them are currently fragile ($n \leq m$) (the remaining $n - m$ sites are currently solid). The dying regions are randomly selected from $n$ currently fragile regions, while the newly born regions are randomly selected from $m - n$ solid regions.

The simplest TFBM with a fixed rate of the "birth and death" process is defined by the parameters $m, n$, and *turnover* rate $x$ (FBM is a particular case of TFBM corresponding to $x = 0$, while RBM is a particular case of TFBM corresponding to $x = 0$ and $n = m$). While this over-simplistic model with a fixed turnover rate may not adequately describe the real rearrangement process, it allows one to analyze the general trends and to compare them to the trends observed in real data.

The leftmost subtable of Table 4 with $x = 0$ represents an equivalent of Table 1(left panel) for FBM and reveals that the inter-reuse is roughly the same on all pairs of branches ($\approx 110$ for $n = 500$, $\approx 70$ for $n = 900$, $\approx 50$ for $n = 1300$). The right subtables of Table 4 represent equivalents of the leftmost subtable for TFBM with the turnover rate $x = 1, 2, 3$ and reveal that the inter-reuse in yellow cells is lower than in green cells, while the inter-reuse in green cells is lower than in red cells.

Fig. 6 shows the scaled inter-reuse averaged over yellow, green, and red cells that reveals a different behavior between FBM and TFBM. Indeed, while the scaled inter-reuse is close to 1 for all pairs of branches in the case of FBM, it varies in the case of TFBM. For example, for $n = 900, m = 2000$, and $x = 3$, the inter-reuse in yellow cells is $\approx 40$, in green cells is $\approx 45$, and in red cells is $\approx 56$. In the following sections we describe an accurate formula for estimating the breakpoint inter-reuse in the case of TFBM that accurately approximates the values shown in Fig. 6.

Our simulations demonstrate that the distribution of inter-reuses among green, red, and yellow cells differs between FBM and TFBM. We argue that this distribution (e.g., the slope of the curve in Fig. 6) represents yet another test to confirm or reject FBM/TFBM. However, while it is clear how to apply this test to the simulated data (with known rearrangements), it remains unclear how to compute it for real data when the ancestral genomes (as well as the parameters of the model) are unknown. While the ancestral genomes can be reliably approximated using the algorithms for ancestral genome reconstruction [4, 25, 26, 45], estimating the number of fragile regions remains an open problem (see [38]). Below we develop a new test (that does not require knowledge of the number of the fragile regions $n$) and demonstrate that FBM does not pass this test while TFBM does, explaining the surprisingly low inter-reuse in mammalian genomes.

**Multispecies breakpoint reuse test**

Given a phylogenetic tree describing a rearrangement scenario, we define the multispecies breakpoint reuse on this tree as follows. For two rearrangements $\rho_1$ and $\rho_2$ in the scenario, we define the distance $d(\rho_1, \rho_2)$ as the number of rearrangements in the scenario between $\rho_1$ and $\rho_2$ plus 1. For example, the distance between 2-breaks $r_4$ and $r_6$ in the tree in Fig. 3 is 4. We define the (actual) multispecies breakpoint reuse as a function

$$R(\ell) = \frac{\sum_{\rho_1, \rho_2 \,:\, d(\rho_1, \rho_2) = \ell} br(\rho_1, \rho_2)}{\sum_{\rho_1, \rho_2 \,:\, d(\rho_1, \rho_2) = \ell} 1}$$

that represents the total breakpoint reuse between pairs of rearrangements $\rho_1$, $\rho_2$ at the distance $\ell$ divided by the number of such pairs. Here $br(\rho_1, \rho_2)$ stands for the number of vertices used by both 2-breaks $\rho_1$ and $\rho_2$.

Since the rearrangements on branches of the phylogenetic tree are unknown, we use the following sampling procedure to approximate $R(\ell)$. Given genomes $P$ and $Q$, we sample various shortest rearrangement scenarios between these genomes by generating random 2-break transformations of $P$ into $Q$. To generate a random transformation we first randomly select a non-trivial cycle $C$ in the breakpoint graph $G(P, Q)$ with the probability proportional to $|C|/2 - 1$, i.e., the number of 2-breaks required to transform such a cycle into a collection of trivial cycles ($|C|$ stands for the length of $C$). Then we uniformly randomly select a 2-break $\rho$ from the set of all $\binom{|C|/2}{2} = \frac{|C|(|C|-2)}{8}$ 2-breaks that splits the selected cycle $C$ into two and thus by Theorem 1 decreases the distance between $P$ and $Q$ by 1 (i.e., $d(\rho P, Q) = d(P, Q) - 1$). We continue selecting non-trivial cycles and 2-breaks in an iterative fashion for genomes $\rho \cdot P$ and $Q$ and so on until $P$ is transformed into $Q$.

The described sampling can be performed for every branch $e = (P, Q)$ of the phylogenetic tree, essentially partitioning $e$ into $length(e) = d(P, Q)$ sub-branches, each featuring a single 2-break. The resulting tree will have $\sum_e length(e)$ sub-branches, where the sum is taken over all branches $e$.

For each pair of sub-branches, we compute the number of reused vertices across them and accumulate these numbers according to the distance between these sub-branches in the tree. The *empirical multispecies breakpoint reuse* (the average reuse between all sub-branches at the distance $\ell$) is defined as the actual multispecies breakpoint reuse in a sampled rearrangement scenario. Our tests on phylogenetic trees with varying topologies demonstrated a good fit between the actual, empirical, and theoretical $R(l)$ curves (data are not shown).

For the five mammalian genomes, the plot of $R(\ell)$ is shown in Fig. 7. From this empirical curve we estimated the parameters $n \approx 196$, $x \approx 1.12$, and $m \approx 4017$ (see the next section) and displayed the corresponding theoretical curve. We remark that the estimated parameter $n$ in TFBM is expected to be larger than the observed number of synteny blocks (since not all potentially breakable regions were broken in a given evolutionary scenario).

We argue that the empirical multispecies breakpoint reuse curve $R(\ell)$ complements the "exponential length distribution" [34] and "pairwise breakpoint reuse" [38] tests as the 3rd criterion to accept/reject RBM, FBM, and now TFBM. One can use the parameters $n$ and $x$ (estimated from empirical $R(\ell)$ curve) to evaluate the extent of the "birth and death" process and to explain why Ma et al., 2006 [25] found so few shared breakpoints between different mammalian lineages. In practice, the "multispecies breakpoint reuse test" can be applied in the same way as the Nadeau-Taylor "exponential length distribution test" was applied in numerous papers. The Nadeau-Taylor test typically amounted to constructing a histogram of synteny blocks and evaluating (often visually) whether it fits the exponential distribution. Similarly, the "multispecies breakpoint reuse test" amounts to constructing $R(\ell)$ curve and evaluating whether it significantly deviates from a horizontal line suggested by RBM and FBM. The estimated parameters of the TFBM model (see the next section) can be used to quantify the extent of these deviations.

TFBM also raises an intriguing question of what triggers the birth and death of fragile regions. As demonstrated by Zhao and Bourque, 2009 [50], the disproportionately large number of rearrangements in primate lineages are flanked by MSDs. TFBM is consistent with the Zhao-Bourque hypothesis that rearrangements are triggered by MSDs since MSDs are also subject to the "birth and death" process. Indeed, after a segmental duplication the pair of matching segments becomes subjected to random mutations and the similarity between these segments dissolves with time (a pair of segmental duplications "disappears" after $\approx$ 40 million years of evolution if one adopts the parameters for defining segmental duplications from [15]). The mosaic structure of segmental duplications [15] provides an additional explanation of how MSDs may promote breakpoint re-uses and generate long cycles typical for the breakpoint graphs of mammalian genomes.

**Computing multispecies breakpoint reuse in the TFBM model**

Let *Fragile* and *Solid* be the sets of $n$ initial fragile regions and $m - n$ initial solid regions respectively. In TFBM, the sets *Fragile* and *Solid* change in accordance with the turnover rate $x$, i.e., after every 2-break $x$ randomly chosen regions (corresponding to $2x$ vertices in the breakpoint graph) from *Fragile* are moved to *Solid*, and vice versa. For a vertex in the set *Fragile*, we evaluate the probability $P(\ell)$ that this vertex still belongs to *Fragile* after $\ell$ 2-breaks. After every 2-break, a vertex from *Fragile* moves to *Solid* with the probability $\frac{x}{n}$, while a vertex from *Solid* moves to *Fragile* with the probability $\frac{x}{m-n}$. Therefore,

$$P(\ell+1) = P(\ell) \cdot \left(1 - \frac{x}{n}\right) + (1 - P(\ell)) \cdot \frac{x}{m-n} = \left(1 - \frac{xm}{n(m-n)}\right) \cdot P(\ell) + \frac{x}{m-n}$$

with $P(0) = 1$. Solution to this recurrence is $P(\ell) = \frac{m-n}{m}\left(1 - \frac{xm}{n(m-n)}\right)^\ell + \frac{n}{m}$. We now compute the expected reuse between 2-breaks $\rho_1$ and $\rho_2$ separated by $\ell$ other 2-breaks. Since every 2-break uses 4 vertices, the probability that it uses a particular vertex in *Fragile* is $\frac{2}{n}$. Since the 2-break $\rho_1$ used 4 vertices, the expected

reuse between $\rho_1$ and $\rho_2$ is:

$$R(\ell) = 4 \cdot \frac{2}{n} \cdot P(\ell) = \frac{8 \cdot (m-n)}{n \cdot m} \left(1 - \frac{xm}{n(m-n)}\right)^\ell + \frac{8}{m}.$$

This formula fits the simulated data well, thus opening a possibility to determine the parameters $m$, $n$, and $x$ for given real genomes. In particular, $n$ and $x$ can be determined from the value and slope of $R(\ell)$ at $\ell = 0$, since $R(0) = \frac{8}{n}$ and $R'(0) \approx -\frac{8x}{n^2}$ (assuming $\frac{xm}{n(m-n)} \ll 1$).

**Fragile regions in the human genome**

Let us imagine the following gedanken experiment: 25 million years ago (time of the human-macaque split) a scientist sequences the genome of the human-macaque ancestor (*QH*) and attempts to predict the sites of (future) rearrangements in the (future) human genome. The only other information the scientist has is the mouse, rat, and dog genomes. While RBM offers no clues on how to make such a prediction, FBM suggests that the scientist should use the breakpoints between one of the available genomes and *QH* as a proxy for fragile regions. For example, there are 552 breakpoints between the mouse genome (*M*) and *QH* and 34 of them were actually used in the human lineage, resulting in only $34/552 \approx 6\%$ accuracy in predicting future human breakpoints (we use synteny blocks larger than 500K from [4]).

TFBM suggests that the scientist should rather use the *closest* genome to *QH* to better predict the human breakpoints. That can be achieved by first reconstructing the common ancestor (*MRD*) of mouse, rat, dog, and human-macaque ancestor and then using the breakpoints between *MRD* and *QH* as a proxy for the sites of rearrangements in the human lineage. 18 out 162 breakpoints between *MRD* and *QH* were used in the human lineage, resulting in $18/162 \approx 11\%$ accurate prediction of human breakpoints, nearly doubling the accuracy of predictions from distant genomes.

Now let us imagine that the scientist somehow gained access to the extant macaque genome. There are 68 breakpoints between *Q* and *QH* and 10 of them were used in the human lineage, resulting in $10/68 \approx 16\%$ accurate prediction of human breakpoints, again improving the accuracy of predictions.

These estimates indicate that TFBM can be used to improve the prediction accuracy of *future* rearrangements in various lineages and demonstrate that the sites of *recent* rearrangements in the human and other primate lineages represent the best guess for the currently active fragile regions in the human genome.

We therefore focus on the incident branches *H+*, *Q+*, and *QH+* and construct the breakpoint graphs $G(H, QH)$, $G(Q, QH)$, and $G(QH, MRD)$. We further superimposed these three graphs to find out breakpoints that were inter-reused on the branches *H+*, *Q+*, and *QH+*. Figure 8 shows the positions of these recently affected breakpoints (projected to the human genome) that, according to TFBM, represent the best proxy for the currently active fragile regions in the human genome. Various ongoing primate genome sequencing projects will soon result in an even better estimate for the fragile regions in the human genome.

## Discussion

Since every species on Earth (including *Homo sapiens*) may speciate into multiple new species, one can ask a question: "How will the human genome evolve in the *next* million years?" TFBM suggests the putative sites of *future* rearrangements in the human genome. The answer to the question "Where are the (future) fragile regions in the human genome?" may be surprisingly simple: they are likely to be among the breakpoint regions that were used in various primate lineages.

Nadeau and Taylor, 1984 [34] proposed RBM based on a single observation: the exponential distribution of the human-mouse synteny block sizes. There is no doubt that jumping to this conclusion was not fully justified: there are many other models (e.g., FBM) that lead to the same exponential distribution of the "visible" synteny block sizes. Currently, there is no single piece of evidence that would allow one to claim that RBM is correct and FBM is not.

While Pevzner and Tesler, 2003 [38] revealed large breakpoint reuse (supporting FBM and contradicting RBM), Ma et al., 2006 [25] discovered low breakpoint inter-reuse (contradicting FBM), calling for yet another generalization of FBM. The proposed TFBM model not only passes both "exponential length distribution" test (motivation for RBM) and "pairwise breakpoint reuse" test (motivation for FBM) but also explains the puzzling discovery of limited breakpoint inter-reuse in [25]. We therefore argue that TFBM is a more accurate model of chromosome evolution, allowing one to approximate the currently active fragile regions in the human genome.

Needless to say, TFBM, similarly to RBM and FBM (or various models of point mutations, e.g., Jukes-Cantor model), is a simplistic model of chromosome evolution that is only an approximation of the real evolutionary process. Moreover, in the current paper we considered TFBM only for the case of 2-breaks and did not include other rearrangements such as transpositions. However, it is fair to assume that transpositions are as likely to happen on incident branches as on distant branches, implying that they cannot possibly cause the reduced breakpoint inter-reuse on distant branches. In addition to limitations of TFBM as a model, there exists a concern whether computation of empirical multispecies breakpoint reuse (that requires reconstruction of ancestral genomes) may be affected by errors in reconstruction of ancestral genomes. While various tools for ancestral genome reconstruction (such as MGRA [4] and inferCARs [25]) were shown to be quite accurate (in particular, they produce nearly identical results while using very different algorithms), it is a challenging open problem to evaluate the multispecies breakpoint reuse without explicitly computing ancestral genomes.

The key point of this paper is the birth and death process of fragile regions rather than a specific model aimed at estimating the hidden parameters of this process. TFBM is merely an initial and over-simplistic attempt to estimate these parameters. The parameters predicted by TFBM (e.g., the number of active fragile regions) are currently difficult to superimpose with scarce information about rearrangements in only 7 reliably completed mammalian genomes, not unlike the parameters of RBM derived in 1984 when no high-resolution compar-ative mammalian genomic architectures were available. However, similarly to comparative mapping efforts in early 1990s that confirmed the Nadeau-Taylor estimates, we believe that imminent sequencing of over 400 primate species will soon provide the detailed information about chromosomal fragility in human genome and will allow one to verify the TFBM parameters.

Similarly to the discovery of breakpoint reuse in 2003 [38], there is currently only indirect evidence supporting the birth and death of fragile regions in chromosome evolution. However, we hope that, similarly to FBM (that led to many follow-up studies supporting the existence of fragile regions), TFBM will trigger further investigations of the fragile regions longevity.

## Acknowledgements

The authors thank Glenn Tesler and Jian Ma for many helpful comments.

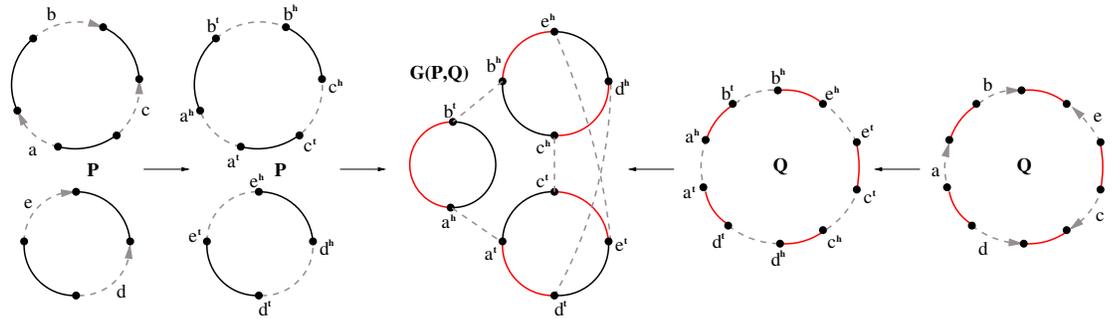

Fig. 1: The breakpoint graph $G(P, Q)$ of a two-chromosomal genome $P = (+a+b-c)(-d+e)$ and a unichromosomal genome $Q = (+a+b-e+c-d)$ represented as two black-obverse cycles and a red-obverse cycle correspondingly. The directions of obverse edges (shown as dashed edges) are not shown in the cases when they are defined by superscripts "$t$" and "$h$".

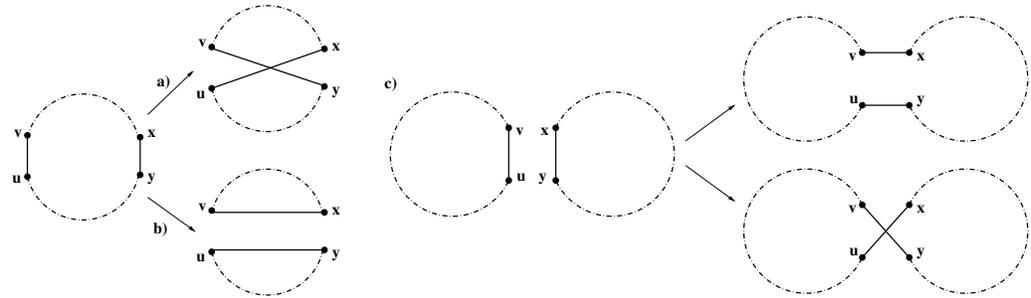

Fig. 2: A 2-break on edges $(u, v)$ and $(x, y)$ corresponding to a) reversal; b) fission; c) translocation/fusion.

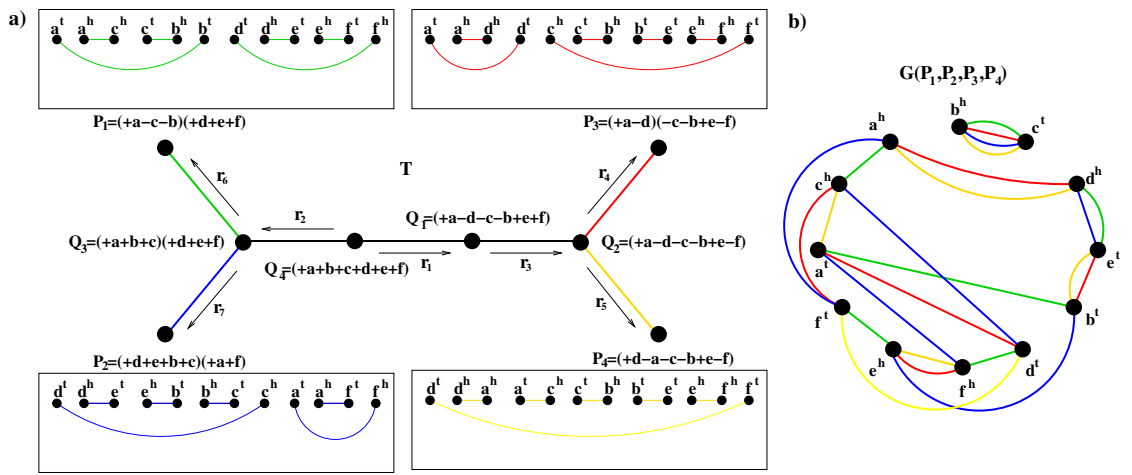

Fig. 3: a) A phylogenetic tree with four circular genomes $P_1, P_2, P_3, P_4$ (represented as green, blue, red, and yellow graphs respectively) at the leaves and specified intermediate genomes. The obverse edges are not shown. b) The multiple breakpoint graph $G(P_1, P_2, P_3, P_4)$ is a superposition of graphs representing genomes $P_1, P_2, P_3, P_4$.

|        | branch lengths = 100 |||||||  branch lengths as in Fig. 4 |||||||
|--------|-----|-----|-----|-----|-----|-----|-----|-----|-----|-----|-----|-----|-----|-----|
| $n = 500$ | M+ | R+ | D+ | Q+ | H+ | MR+ | QH+ | M+ | R+ | D+ | Q+ | H+ | MR+ | QH+ |
| M+  | 63:70 | 106:106 | 103:103 | 97:97 | 108:108 | 98:98 | 113:113 | 23 | 48 | 71 | 16 | 22 | 99 | 41 |
| R+  |       | 57:70   | 103:103 | 108:108 | 98:98 | 102:102 | 122:122 |    | 34 | 83 | 19 | 25 | 116 | 49 |
| D+  |       |         | 65:74   | 104:104 | 125:125 | 104:104 | 106:106 |    |    | 78 | 26 | 37 | 171 | 74 |
| Q+  |       |         |         | 58:68   | 126:126 | 120:120 | 120:120 |    |    |    | 2  | 9  | 39  | 16 |
| H+  |       |         |         |         | 56:62   | 113:113 | 116:116 |    |    |    |    | 6  | 51  | 22 |
| MR+ |       |         |         |         |         | 71:84   | 104:104 |    |    |    |    |    | 186 | 102 |
| QH+ |       |         |         |         |         |         | 54:60   |    |    |    |    |    |     | 25 |
| $n = 900$ | M+ | R+ | D+ | Q+ | H+ | MR+ | QH+ | M+ | R+ | D+ | Q+ | H+ | MR+ | QH+ |
| M+  | 37:38 | 70:70 | 83:83 | 90:90 | 72:72 | 76:76 | 87:87 | 13 | 30 | 44 | 9  | 13 | 67 | 25 |
| R+  |       | 47:50 | 67:67 | 63:63 | 74:74 | 68:68 | 49:49 |    | 20 | 53 | 11 | 16 | 79 | 31 |
| D+  |       |       | 37:38 | 69:69 | 62:62 | 78:78 | 84:84 |    |    | 46 | 17 | 24 | 121 | 45 |
| Q+  |       |       |       | 32:36 | 76:76 | 75:75 | 94:94 |    |    |    | 1  | 4  | 24 | 9  |
| H+  |       |       |       |       | 40:44 | 64:64 | 68:68 |    |    |    |    | 4  | 34 | 13 |
| MR+ |       |       |       |       |       | 42:44 | 64:64 |    |    |    |    |    | 113 | 70 |
| QH+ |       |       |       |       |       |       | 28:28 |    |    |    |    |    |     | 14 |
| $n = 1300$ | M+ | R+ | D+ | Q+ | H+ | MR+ | QH+ | M+ | R+ | D+ | Q+ | H+ | MR+ | QH+ |
| M+  | 42:46 | 46:46 | 52:52 | 51:51 | 47:47 | 62:62 | 39:39 | 8 | 21 | 33 | 7 | 9 | 52 | 19 |
| R+  |       | 31:34 | 53:53 | 66:66 | 54:54 | 48:48 | 56:56 |   | 13 | 39 | 8 | 11 | 60 | 24 |
| D+  |       |       | 25:26 | 64:64 | 62:62 | 60:60 | 64:64 |   |    | 34 | 12 | 17 | 91 | 34 |
| Q+  |       |       |       | 22:22 | 58:58 | 50:50 | 50:50 |   |    |    | 1  | 3  | 19 | 7 |
| H+  |       |       |       |       | 30:30 | 57:57 | 72:72 |   |    |    |    | 2  | 25 | 10 |
| MR+ |       |       |       |       |       | 31:34 | 42:42 |   |    |    |    |    | 81 | 51 |
| QH+ |       |       |       |       |       |       | 19:20 |   |    |    |    |    |    | 9 |

Table 1: **Left panel:** The number of intra- and inter-reuses between 7 branches of the tree in Fig. 4, each of length 100, for simulated genomes with $n$ fragile regions ($n$ = 500, 900, 1300). The diagonal elements represent intra-reuses while the elements above diagonal represent inter-reuses. In each cell with numbers $x : y$, $x$ represents the observed reuse while $y$ represents the corresponding lower bound. The cells of the table are colored red (for adjacent branches like $M+$ and $R+$), green (for branches that are separated by a single branch like $M+$ and $D+$ separated by $MR+$), and yellow (for branches that are separated by two branches like $M+$ and $H+$ separated by $MR+$ and $QH+$). **Right panel:** The estimated number of intra- and inter-reuses $bound(e)$ and $bound(e_1, e_2)$ between 7 branches with varying branch lengths as specified in Fig. 4 (data simulated according to FBM with $n$ fragile regions).

|     | M+ | R+ | D+ | Q+ | H+ | MR+ | QH+ |
|-----|----|----|----|----|----|-----|-----|
| M+  | 84 | 68 | 20 | 4  | 5  | 58  | 15  |
| R+  |    | 96 | 22 | 3  | 6  | 60  | 17  |
| D+  |    |    | 174 | 17 | 19 | 98  | 64  |
| Q+  |    |    |    | 12 | 10 | 25  | 18  |
| H+  |    |    |    |    | 22 | 23  | 18  |
| MR+ |    |    |    |    |    | 292 | 80  |
| QH+ |    |    |    |    |    |     | 70  |

Table 2: The estimated number of intra- and inter-reuses $bound(e)$ and $bound(e_1, e_2)$ between 7 branches of the phylogenetic tree in Fig. 4 of five mammalian genomes (real data).

|     | M+ | R+ | D+ | Q+ | H+ | MR+ | QH+ |
|-----|----|----|----|----|----|-----|-----|
| M+  | 13 | 30 | 44 | 9  | 13 | 67  | 25  |
| R+  | 30 | 20 | 53 | 11 | 16 | 79  | 31  |
| QH+ | 25 | 31 | 45 | 9  | 13 | 70  | 14  |
| M+  | 84 | 68 | 20 | 4  | 5  | 58  | 15  |
| R+  | 68 | 96 | 22 | 3  | 6  | 60  | 17  |
| QH+ | 15 | 17 | 64 | 18 | 18 | 80  | 70  |

Table 3: Subtables of Table 1(right panel) for $n$ = 900 (top part) and Table 2 (bottom part) featuring branches $M+$, $R+$, and $QH+$ as one element of the pair.

|       | $x = 0$ (FBM) |||||||  $x = 1$ |||||||  $x = 2$ |||||||  $x = 3$ |||||||
|-------|----|-----|-----|-----|-----|-----|-----|----|----|----|----|----|-----|-----|----|----|----|----|----|-----|-----|----|----|----|----|----|-----|-----|
| $n = 500$ | M+ | R+ | D+ | Q+ | H+ | MR+ | QH+ | M+ | R+ | D+ | Q+ | H+ | MR+ | QH+ | M+ | R+ | D+ | Q+ | H+ | MR+ | QH+ | M+ | R+ | D+ | Q+ | H+ | MR+ | QH+ |
| M+  | 67 | 110 | 109 | 109 | 110 | 111 | 108 | 64 | 93 | 75 | 66 | 65 | 92 | 77 | 63 | 78 | 57 | 45 | 46 | 79 | 57 | 58 | 69 | 47 | 36 | 36 | 68 | 46 |
| R+  |    | 69  | 110 | 110 | 108 | 109 | 107 |    | 67 | 76 | 65 | 65 | 92 | 78 |    | 63 | 57 | 46 | 45 | 78 | 58 |    | 60 | 47 | 36 | 37 | 69 | 46 |
| D+  |    |     | 69  | 109 | 108 | 109 | 109 |    |    | 66 | 76 | 77 | 91 | 90 |    |    | 62 | 56 | 58 | 78 | 77 |    |    | 58 | 46 | 47 | 68 | 69 |
| Q+  |    |     |     | 68  | 108 | 109 | 110 |    |    |    | 65 | 92 | 77 | 93 |    |    |    | 61 | 79 | 57 | 76 |    |    |    | 60 | 68 | 48 | 66 |
| H+  |    |     |     |     | 71  | 107 | 109 |    |    |    |    | 66 | 78 | 94 |    |    |    |    | 61 | 58 | 79 |    |    |    |    | 60 | 48 | 67 |
| MR+ |    |     |     |     |     | 70  | 109 |    |    |    |    |    | 65 | 91 |    |    |    |    |    | 62 | 77 |    |    |    |    |    | 57 | 69 |
| QH+ |    |     |     |     |     |     | 68  |    |    |    |    |    |    | 65 |    |    |    |    |    |    | 61 |    |    |    |    |    |    | 59 |
| $n = 900$ | M+ | R+ | D+ | Q+ | H+ | MR+ | QH+ | M+ | R+ | D+ | Q+ | H+ | MR+ | QH+ | M+ | R+ | D+ | Q+ | H+ | MR+ | QH+ | M+ | R+ | D+ | Q+ | H+ | MR+ | QH+ |
| M+  | 42 | 71 | 72 | 71 | 71 | 71 | 71 | 40 | 64 | 58 | 53 | 54 | 60 | 60 | 39 | 60 | 51 | 45 | 45 | 59 | 51 | 37 | 55 | 45 | 39 | 39 | 56 | 44 |
| R+  |    | 41 | 72 | 71 | 72 | 72 | 73 |    | 39 | 59 | 53 | 54 | 65 | 58 |    | 38 | 51 | 45 | 45 | 61 | 51 |    | 38 | 45 | 39 | 39 | 56 | 45 |
| D+  |    |    | 40 | 73 | 72 | 70 | 72 |    |    | 41 | 60 | 59 | 65 | 65 |    |    | 38 | 50 | 51 | 58 | 60 |    |    | 38 | 46 | 46 | 56 | 54 |
| Q+  |    |    |    | 39 | 73 | 71 | 73 |    |    |    | 39 | 64 | 58 | 64 |    |    |    | 38 | 59 | 49 | 60 |    |    |    | 37 | 55 | 46 | 56 |
| H+  |    |    |    |    | 41 | 71 | 71 |    |    |    |    | 38 | 59 | 64 |    |    |    |    | 38 | 49 | 59 |    |    |    |    | 37 | 46 | 55 |
| MR+ |    |    |    |    |    | 40 | 74 |    |    |    |    |    | 40 | 66 |    |    |    |    |    | 40 | 61 |    |    |    |    |    | 37 | 55 |
| QH+ |    |    |    |    |    |    | 41 |    |    |    |    |    |    | 39 |    |    |    |    |    |    | 40 |    |    |    |    |    |    | 37 |
| $n = 1300$ | M+ | R+ | D+ | Q+ | H+ | MR+ | QH+ | M+ | R+ | D+ | Q+ | H+ | MR+ | QH+ | M+ | R+ | D+ | Q+ | H+ | MR+ | QH+ | M+ | R+ | D+ | Q+ | H+ | MR+ | QH+ |
| M+  | 28 | 54 | 52 | 54 | 52 | 53 | 55 | 27 | 48 | 46 | 45 | 44 | 49 | 47 | 27 | 46 | 44 | 40 | 39 | 48 | 41 | 28 | 45 | 40 | 37 | 38 | 45 | 40 |
| R+  |    | 28 | 53 | 53 | 54 | 53 | 52 |    | 29 | 45 | 44 | 44 | 48 | 48 |    | 28 | 43 | 40 | 41 | 46 | 44 |    | 27 | 41 | 38 | 37 | 44 | 39 |
| D+  |    |    | 31 | 52 | 51 | 53 | 54 |    |    | 28 | 46 | 47 | 50 | 49 |    |    | 29 | 42 | 42 | 47 | 46 |    |    | 27 | 39 | 41 | 44 | 46 |
| Q+  |    |    |    | 28 | 52 | 55 | 53 |    |    |    | 29 | 50 | 46 | 50 |    |    |    | 28 | 49 | 42 | 47 |    |    |    | 27 | 46 | 39 | 45 |
| H+  |    |    |    |    | 29 | 53 | 52 |    |    |    |    | 28 | 47 | 49 |    |    |    |    | 27 | 42 | 46 |    |    |    |    | 27 | 41 | 44 |
| MR+ |    |    |    |    |    | 27 | 53 |    |    |    |    |    | 29 | 49 |    |    |    |    |    | 27 | 48 |    |    |    |    |    | 27 | 46 |
| QH+ |    |    |    |    |    |    | 29 |    |    |    |    |    |    | 28 |    |    |    |    |    |    | 28 |    |    |    |    |    |    | 27 |

Table 4: The breakpoint intra- and inter-reuse (averaged over 100 simulations) for five simulated genomes $M, R, D, Q, H$ under TFBM model with $m = 2000$ synteny blocks, $n$ fragile regions, the turnover rate $x$, and the evolutionary tree shown in Fig. 4 with the length of each branch equal 100.

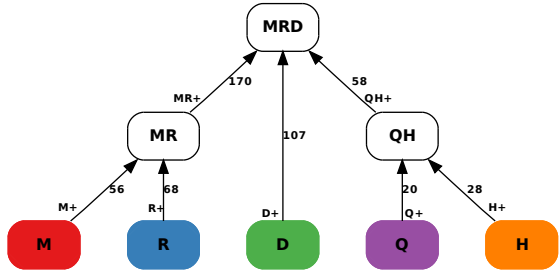

Fig. 4: The phylogenetic tree $T$ on five genomes $M, R, D, Q,$ and $H$. The branches of the tree are denoted as $M+, R+, D+, Q+, H+, MR+,$ and $QH+$.

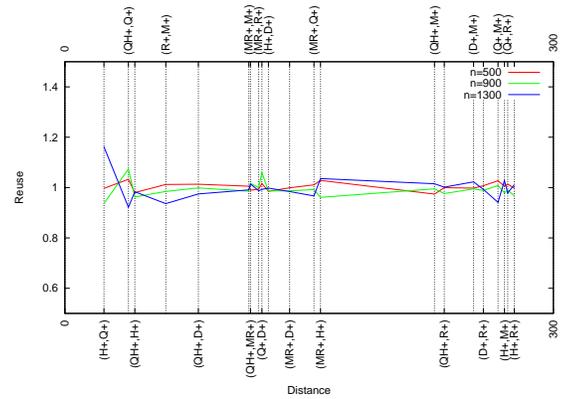

Fig. 5: The scaled inter-reuse for five simulated genomes $M, R, D, Q, H$ (averaged over 100 simulations) on $n$ fragile regions (for $n = 500, 900,$ and $1300$) with the evolutionary tree and branch lengths shown in Fig. 4.

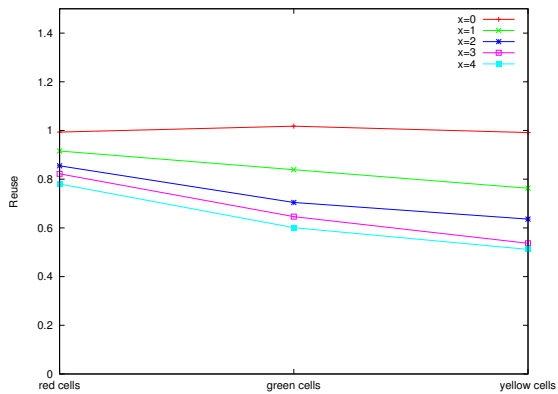

Fig. 6: The scaled inter-reuse for five simulated genomes $M, R, D, Q, H$ on $m = 2000$ synteny blocks, $n = 900$ fragile regions, and the turnover rate $x$ varying from 0 to 4 with the phylogenetic tree and branch lengths shown in Fig. 4. The simulations follow FBM ($x = 0$) and TFBM ($x$ varies from 1 to 4). The plot shows the scaled inter-reuse for only three reference points (corresponding to red, green, and yellow cells) that are somewhat arbitrarily connected by straight segments for better visualization.

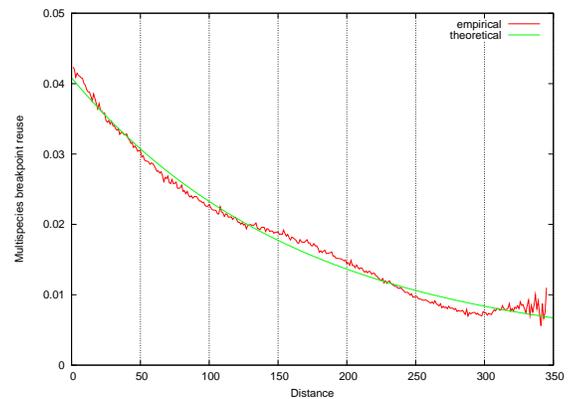

Fig. 7: Empirical and theoretical curves representing the number of reuses $R(\ell)$ as a function of distance $\ell$ between pairs of sub-branches of the tree in Fig. 4 of the five mammalian genomes (ancestral genomes were computed using MGRA [4]). The empirical curve is averaged over 1000 random samplings of shortest rearrangement scenarios, while the theoretical curve represents the best fit with parameters $n \approx 196$, $x \approx 1.12$, and $m \approx 4017$.

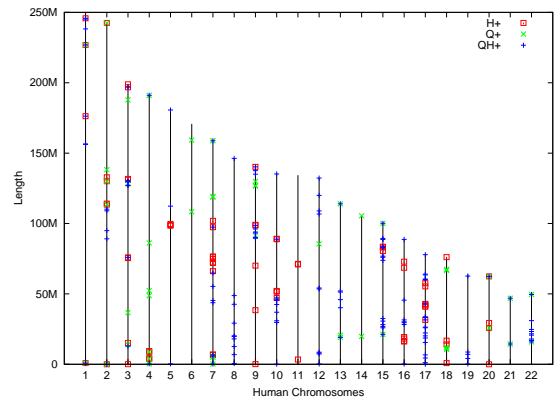

Fig. 8: Positions of regions broken on the evolutionary path from the rodent-primate-carnivore ancestor (i.e., on $H+$, $Q+$, and $QH+$ branches) projected to the human chromosomes.